\documentclass[aps,prc,showpacs,showkeys,twoside,twocolumn,byrevtex,floatfix]{revtex4-2}
\usepackage{graphicx}        % PRC 
\usepackage{color}           % PRC % for color  usage: \textcolor{red}{text}
\usepackage{amsmath,amssymb,amsfonts}
\usepackage{bm}
\usepackage{mathtools}
\usepackage{ulem}

\newcommand{\nc}{\newcommand}           % new command
\nc{\vc}[1]     {\mbox{\boldmath $#1$}} % boldmath(vector)
\nc{\mapleft}[1]{                       % something under arrow
 \smash{\mathop{                      %
  \hbox to 0.90cm{\rightarrowfill} }\limits_{#1}}}

\nc{\beq}     {\begin{eqnarray}}
\nc{\eeq}    {\end{eqnarray}}
\nc{\bra}       {\langle}               % bra
\nc{\ket}       {\rangle}               % ket
\nc{\bras}[1]   {\langle#1|}            % <#1|
\nc{\kets}[1]   {|#1\rangle}            % |#1>
\nc{\del}       {\partial}              % bra state

\newcommand{\lw}[1]{\smash{\lower1.75ex\hbox{#1}}}

\nc{\red}[1]    {\textcolor{black}{#1}}  % red
\nc{\mydraft}	{\setlength{\topmargin}{-1.5cm}}
\mydraft
\begin{document}

\title{
  Quadrupole transitions of $^{10}$C and their isospin symmetry with $^{10}$Be
}

\author{Takayuki Myo}
\email{takayuki.myo@oit.ac.jp}
\affiliation{General Education, Faculty of Engineering, Osaka Institute of Technology, Osaka, Osaka 535-8585, Japan}
\affiliation{Research Center for Nuclear Physics (RCNP), Osaka University, Ibaraki, Osaka 567-0047, Japan}

\author{Mengjiao Lyu}
\affiliation{College of Science, Nanjing University of Aeronautics and Astronautics, Nanjing 210016, China}

\author{Qing Zhao}
\affiliation{School of Science, Huzhou University, Huzhou 313000, Zhejiang, China}

\author{Masahiro Isaka}
\affiliation{Faculty of Intercultural Communication, Hosei University, Chiyoda-ku, Tokyo 102-8160, Japan}

\author{Niu Wan}
\affiliation{School of Physics and Optoelectronics, South China University of Technology, Guangzhou 510641, China}

\author{Hiroki Takemoto}
\affiliation{Faculty of Pharmacy, Osaka Medical and Pharmaceutical University, Takatsuki, Osaka 569-1094, Japan} 
\affiliation{Research Center for Nuclear Physics (RCNP), Osaka University, Ibaraki, Osaka 567-0047, Japan}

\author{Hisashi Horiuchi}
\affiliation{Research Center for Nuclear Physics (RCNP), Osaka University, Ibaraki, Osaka 567-0047, Japan}

\author{Akinobu Dot\'e}
\affiliation{KEK Theory Center, Institute of Particle and Nuclear Studies (IPNS), High Energy Accelerator Research Organization (KEK), Tsukuba, Ibaraki, 305-0801, Japan}
\affiliation{J-PARC Branch, KEK Theory Center, IPNS, KEK, Tokai, Ibaraki, 319-1106, Japan}
\affiliation{Graduate Institute for Advanced Studies, SOKENDAI, 1-1 Oho, Tsukuba, Ibaraki, 305-0801, Japan} % add 2025.09.04

\author{Hiroshi Toki}
\affiliation{Research Center for Nuclear Physics (RCNP), Osaka University, Ibaraki, Osaka 567-0047, Japan}

\date{\today}

\begin{abstract}%
  We investigate the structures of $^{10}$C focusing on the quadrupole properties in comparison with the mirror nucleus $^{10}$Be.
  We describe $^{10}$C and $^{10}$Be in the variation of the multiple bases of the antisymmetrized molecular dynamics (AMD),
  in which the multiple AMD bases are optimized simultaneously in the total-energy variation.
  In the monopole transitions, we confirm the isospin symmetry between $^{10}$C and $^{10}$Be by exchanging protons and neutrons.
  In the quadrupole transitions, most cases show larger values in $^{10}$C than those of $^{10}$Be, except for the transition of $2^+_1\to 0^+_1$.
  The transition of $2^+_1\to 0^+_1$ shows similar values in the two nuclei in spite of the different proton numbers,
  which agrees with the experimental situation as an anomaly.
  This relation comes from the small proton deformation in $^{10}$C due to its subclosed nature and the large proton deformation in $^{10}$Be
  due to two-$\alpha$ clustering.
  This property can also be seen in the quadrupole moments of the two nuclei.
  In the neutron deformations of $^{10}$C and $^{10}$Be, the opposite tendency of protons is confirmed and
  these results ensure the isospin symmetry between the two nuclei.
  We also confirm the large quadrupole transitions between the elongated linear-chain states.
  It would be desirable for future experiments to investigate the present characteristics of the transitions in the two nuclei.
\end{abstract}

%\pacs{
%21.60.Gx, % Cluster Models
%}
\maketitle

%%%%%%%%%%%%%%%%%%%%%%%%%%%%%
\section{Introduction}

Nuclear clustering is known as one of the major properties of nuclei \cite{ikeda68,horiuchi12,freer18}.
In the cluster states, some of the nucleons in nuclei form clusters, such as an $\alpha$ particle, and they are spatially developed in nuclei.
A typical case is $^8$Be, which decays into two $\alpha$ particles.
The cluster states are often observed near the threshold energy of the emission of clusters because of the weak interaction between the isolated clusters.
This property is called the ``threshold rule'' \cite{ikeda68}.
In unstable nuclei having excess nucleons,
threshold energies of the emission of valence nucleons/clusters often exist in the low-excitation energy region.
It is interesting to discuss the threshold rule in unstable nuclei in relation to the low-lying threshold energies.

\red{In the nuclear structure including clustering, the electric quadrupole transition strength, $B(E2)$, is a fundamental quantity
  to get knowledge of the deformation and collectivity of protons in nuclei.}
In unstable nuclei, it is interesting to investigate the effect of the valence protons on the nuclear structure, in particular, the deformation property, via $B(E2)$.

In this paper, we investigate a proton-rich nucleus $^{10}$C consisting of $^8$Be and two excess protons,
and compare its structures with those of the mirror system of $^{10}$Be.
\red{In $^{10}$C, the experimental value of the electric quadrupole transition $B(E2,2^+_1\to 0^+_1)$ is 8.8(3) $e^2$fm$^4$,
which is slightly smaller than that of $^{10}$Be at 9.2(3) $e^2$fm$^4$ \cite{mccutchan12}.
Naively, the $B(E2)$ values of the proton-rich nuclei are expected to be larger than that of neutron-rich nuclei in mirror systems.}
This relation is well confirmed in many experimental data of light mirror nuclei \cite{tunl,nndc} in the quadrupole transitions to their ground states,
but the case of $^{10}$C and $^{10}$Be does not follow this relation and becomes the exceptional case.
\red{
To understand the origin of this anomaly in $^{10}$C and $^{10}$Be, we perform a detailed theoretical study of their quadrupole transition properties,
which is the main purpose of this paper.}

For the theoretical method to describe $^{10}$C and $^{10}$Be, we employ the antisymmetrized molecular dynamics (AMD) \cite{kanada03},
because several thresholds of the cluster emissions exist in the low-excitation energy region of the two nuclei \cite{tunl,tilley04}.
In fact, in $^{10}$C, the threshold positions of $^9$B+$p$, $^8$Be+$p$+$p$, $^6$Be+$\alpha$, and $\alpha$+$\alpha$+$p$+$p$ are 4.0 MeV, 3.8 MeV, 5.1 MeV, and 3.7 MeV, respectively.
In relation to the threshold rule, the mean-field state and cluster state can coexist in the two nuclei.
For this purpose, AMD is an appropriate nuclear model to describe various cluster and mean-field structures appearing in a nucleus unifiedly.

In AMD, various configurations are superposed, where the nuclear deformation is treated as the typical constraint to generate the different configurations \cite{horiuchi12,kanada03,suhara10}.
However, it is not trivial whether the configurations obtained with this constraint are the optimal ones to be superposed.
Conversely, we have recently developed a new scheme to optimize the multiple AMD configurations
with respect to the energy variation of the total system without assuming any physical constraints  \cite{myo23b,myo25a}.
We further extend the method to generate the excited-state configurations imposing the orthogonal condition on the ground-state configurations.
The variation of multiple configurations is an advantage of this study because we can optimize many configurations simultaneously without the need for a priori knowledge.
In particular, this scheme is beneficial for constructing the configurations for the excited states.
In the previous works \cite{myo23b,myo25a,tian24,cheng24,tian25}, we have applied this method to light nuclei and hypernuclei, and
have discussed the emergence of the various cluster states together with the mean-field states.
In fact, for $^{10}$Be \cite{myo23b}, among the nuclear cluster models that utilize the same Hamiltonian,
  the present method provides the lowest energies of the ground and excited states. This indicates the advantage of the method from the viewpoint of variational problem.

In AMD, the energy variation is called the cooling method (the imaginary-time evolution) \cite{kanada03}.
We extend the cooling method to the multiple AMD bases and call this method the multiple cooling or ``multicool method''.
In this study, we apply the multicool method to AMD and discuss the structures of $^{10}$C and $^{10}$Be
with the transitions of monopole and quadrupole.
From the results, we discuss the isospin symmetry in $^{10}$C and $^{10}$Be. 

In Sec.~\ref{sec:method}, we explain the formulation of the variation of the multiple AMD configurations.
In Sec.~\ref{sec:result}, we discuss the results of $^{10}$C and $^{10}$Be. 
In Sec.~\ref{sec:summary}, we summarize this work.

%%%%%%%%%%%%%%%%%%%%%%%%%%%%%
\section{Theoretical methods}\label{sec:method}

%%%%%%%%%%%%%%%%%%%%%%%%%%%%%
\subsection{Variation of multiple bases of antisymmetrized molecular dynamics}\label{sec:AMD}

Nuclear wave function of AMD, $\Phi_{\rm AMD}$ is defined as a Slater determinant of an $A$-nucleon system \cite{kanada03}:
\begin{eqnarray}
  \begin{split}
\Phi_{\rm AMD}
&=
\red{
\frac{1}{\sqrt{A!}}
\left|
\begin{array}{cccc}
\phi_1(\bm{r}_1)  & \phi_2(\bm{r}_1) & \cdots & \phi_A(\bm{r}_1) \\
\phi_1(\bm{r}_2)  & \phi_2(\bm{r}_2) & \cdots & \phi_A(\bm{r}_2) \\
\vdots            & \vdots           & \ddots &   \vdots         \\ 
\phi_1(\bm{r}_A)  & \phi_2(\bm{r}_A) & \cdots & \phi_A(\bm{r}_A) \\
\end{array}
\right|}
\\
\phi_i(\bm{r})&=\left(\frac{2\nu}{\pi}\right)^{3/4} e^{-\nu(\bm{r}-\bm{Z}_i)^2} \chi_{\sigma,i} \chi_{\tau,i} ,
\\
\chi_{\sigma,i} &= \alpha^\uparrow_i \kets{\uparrow} + \alpha^\downarrow_i \kets{\downarrow}.
  \end{split}
\label{eq:AMD}
\end{eqnarray}
The single-nucleon wave function $\phi_i(\bm{r})$ with the particle index $i$ has a Gaussian wave packet with a range parameter $\nu$
and the centroid parameter $\bm{Z}_i$ with the condition of $\sum_{i=1}^A \bm{Z}_i={\bf 0}$.
The spin wave function $\chi_{\sigma}$ is a superposition of the up and down components with the weights of $\alpha^{\uparrow/\downarrow}_i$,
which determine the direction of nucleon spin and are treated as the variational parameters in $\Phi_{\rm AMD}$.
The isospin component $\chi_{\tau}$ is a proton or a neutron.
It is noted that the variational parameters $\{\bm{Z}_i,\alpha^{\uparrow/\downarrow}_i\}$ can be complex numbers.

\red{First, we consider the energy variation (energy minimization) of the single AMD basis state $\Phi_{\rm AMD}$.}
The variation is performed in the following cooling equation (imaginary-time evolution) \cite{kanada03}
for total intrinsic energy $E^\pm_{\rm AMD}$ with the parity projection $P^\pm$ for the Hamiltonian $H$, 
\begin{equation}
  \begin{split}
    \Phi^\pm_{\rm AMD}&= P^\pm \Phi_{\rm AMD}  \,, \quad
    E^\pm_{\rm AMD} = \dfrac{ \bra \Phi^\pm_{\rm AMD}|H| \Phi^\pm_{\rm AMD} \ket }{\bra \Phi^\pm_{\rm AMD}| \Phi^\pm_{\rm AMD} \ket},
    \\
    \dfrac{{\rm d} X_i}{{\rm d} t}&= \frac{\mu}{\hbar} \dfrac{\partial E^\pm_{\rm AMD}}{ \partial X_i^*},\quad \mbox{and c.c}.
  \end{split}
  \label{eq:cooling}
\end{equation}
\red{Using an arbitrary negative number $\mu$, the complex parameters $X_i:=\{\bm{Z}_i,\alpha^{\uparrow/\downarrow}_i\}$ are determined.}
After the variation, the angular-momentum projection with the operator $P^J_{MK}$ is performed for the spin state of $(J,M,K)$.
\begin{eqnarray}
  \begin{split}
  \Psi^{J^\pm}_{MK,{\rm AMD}}
  &= P^J_{MK}P^{\pm} \Phi_{\rm AMD}.
  \end{split}
  \label{eq:projection}
\end{eqnarray}
where $J$ and $M$ are the total angular momentum and its $z$-component, respectively, and $K$ is the projection of $J$ onto the intrinsic $z$ axis.

We can superpose $N_{J^\pm}$ different AMD configurations,
where the individual configurations are often generated utilizing the nuclear deformation \cite{horiuchi12,kanada03}.
The total wave function $\Psi_{\rm t}^{J^\pm}$ is a superposition of the projected AMD basis states in Eq.~(\ref{eq:projection}),
denoted as $\Psi_n^{J^\pm}$:
\begin{eqnarray}
  \begin{split}
   \Psi_{\rm t}^{J^\pm}
&= \sum_{n=1}^{N_{J^\pm}} C_n^{J^\pm}\,  \Psi_n^{J^\pm} \,, \quad
   E_{\rm t}^{J^\pm} = \dfrac{ \bra \Psi_{\rm t}^{J^\pm} |H| \Psi_{\rm t}^{J^\pm} \ket }{\bra \Psi_{\rm t}^{J^\pm}| \Psi_{\rm t}^{J^\pm} \ket}, 
  \end{split}
   \label{eq:linear}
\end{eqnarray}
where $n$ the basis index of the projected AMD basis states with $J^\pm$ including the $K$-mixing.
From the variational principle of the total energy, $\delta E_{\rm t}^{J^\pm}=0$, the generalized eigenvalue problem is solved
to obtain $E_{\rm t}^{J^\pm}$ and $\{C_n^{J^\pm}\}$:
\begin{eqnarray}
  \begin{split}
   \sum_{n=1}^{N_{J^\pm}} \Bigl\{ \langle\Psi_{m}^{J^\pm} | H |\Psi_{n}^{J^\pm}\rangle - E_{\rm t}^{J^\pm} \langle\Psi_{m}^{J^\pm} | \Psi_{n}^{J^\pm} \rangle \Bigr\}\, C_n^{J^\pm} &= 0.
  \end{split}
   \label{eq:eigen}
\end{eqnarray}
%
%
%\subsection{Variation of multiple AMD basis states}\label{sec:multi}

Next, we consider the energy variation of the multiple AMD basis states, which is a unique feature of the present study.
\red{We express the total wave function $\Phi$ as a superposition of the intrinsic AMD basis states $\{\Phi_n\}$}
corresponding to $\Phi^\pm_{\rm AMD}$ in Eq. (\ref{eq:cooling}), with a number $N_{\rm b}$ and the total energy $E$ is given as
\begin{equation}
  \begin{split}
   \Phi&= \sum_{n=1}^{N_{\rm b}} C_n\,  \Phi_n , \quad
   E    = \dfrac{ \bra \Phi |H| \Phi \ket }{\bra \Phi | \Phi \ket}.
  \end{split}
  \label{eq:multi}
\end{equation}
Here, the parity projection is always performed and we omit the notation of parity $(\pm)$.

\red{In the previous studies \cite{myo23b,myo25a,tian24,cheng24,tian25},
we have extended the cooling equation for the multiple AMD basis states and we refer to this method simply as the ``multicool method''.}
The properties and many demonstrations of the multicool calculation are given in the above references,
and in this paper, we briefly explain the essential framework of the method;
The multiple AMD configurations have the parameters of $X_{n,i}:=\{\bm{Z}_{n,i},\alpha^{\uparrow/\downarrow}_{n,i},C_n\}$ with the particle index $i$.
The cooling equation is given as
\begin{equation}
  \begin{split}
   \dfrac{{\rm d} X_{n,i}}{{\rm d} t}&= \frac{\mu}{\hbar} \dfrac{\partial E}{ \partial X_{n,i}^*},\quad \mbox{and c.c}.
  \end{split}
  \label{eq:multi_eq}
\end{equation}
This equation is the extension of the one defined in Eq. (\ref{eq:cooling}) by adding the basis index $n$ in the parameters.
From this equation, one determines $\{X_{n,i}\}$ and obtains the ground-state configurations $\{\Phi_n\}$ in Eq.~(\ref{eq:multi}).

\red{
In the next step, to construct the configurations for the excited states, we employ a method that enforces orthogonality to the ground-state configurations $\{\Phi_n\}$.
To achieve this orthogonality, the angular-momentum projection of $\{\Phi_n\}$ is necessary in principle \cite{kanada03},
because $\Phi_n$ can be deformed. However, this procedure is computationally expensive.
As an alternative, we consider an approximated treatment by choosing the several rotations of the configurations of $\{\Phi_n\}$.
In this study, we employ two rotations; $(x,y,z)\to(z,x,y)$ and $(x,y,z)\to(y,z,x)$ as done in the previous works \cite{myo23b,myo25a}.
Each rotation makes $N_{\rm b}$ configurations, and by adding the original $N_{\rm b}$ configurations, $3N_{\rm b}$ configurations are considered in total for the ground state.
We adopt the ground-state configurations $\{\Phi_{c}\}$ with the index $c=1,\cdots,3N_{\rm b}$
and construct the excited-state configurations with the orthogonality to $\{\Phi_{c}\}$.
This method works effectively in the application to the $p$-shell nuclei \cite{myo23b,myo25a}.}

\red{
  Technically, we employ the projection operator method to construct the excited-state configurations \cite{kukulin78}.
  We introduce the pseudopotential $V_\lambda$ using $\{\Phi_{c}\}$ in the projection operator form with a positive strength $\lambda$.
  We define $H_\lambda$ by adding $V_\lambda$ to the Hamiltonian as:}
\begin{equation}
  \begin{split}
    H_\lambda &= H + V_\lambda,\quad
    V_\lambda = \lambda \sum_{c=1}^{3N_{\rm b}} \kets{\Phi_{c}}\bras{\Phi_{c}}.
  \end{split}
  \label{eq:PSE}
\end{equation}
\red{
The expectation value of $V_\lambda$ is positive and has a repulsive effect on the total energy.
In the energy variation with $H_\lambda$, the total wave function is obtained to reduce the expectation value of $V_\lambda$,
namely, the overlap with $\{\Phi_{c}\}$.
If $\lambda$ is sufficiently large, the contribution of $V_\lambda$ to the total energy becomes negligible,
causing the total wave function to become orthogonal to each configuration of $\{\Phi_{c}\}$ \cite{kukulin78}.}
This method has been utilized in the orthogonality condition model of the nuclear cluster systems
to remove the Pauli-forbidden states from the intercluster motions \cite{myo14}.

\red{The total wave function $\Phi_{\lambda}$ and the total energy $E_{\lambda}$ depend on $\lambda$. We write them as}
\begin{equation}
  \begin{split}
    \Phi_{\lambda} &= \sum_{n=1}^{N_{\rm b}} C_{\lambda,n} \Phi_{\lambda,n},\quad
    E_{\lambda} = \dfrac{ \bra \Phi_{\lambda} |H_{\lambda}| \Phi_{\lambda} \ket }{\bra \Phi_{\lambda} | \Phi_{\lambda} \ket}.
  \end{split}
  \label{eq:multi2}
\end{equation}
We perform the variation of $E_{\lambda}$ using Eq. (\ref{eq:multi_eq}) and obtain the configurations $\{\Phi_{\lambda,n}\}$.
We evaluate $E_{\lambda}$ omitting the contribution of $V_\lambda$.

We perform the variation with a small value of $\lambda$ and repeat the calculation increasing $\lambda$.
When $\lambda$ is large, the total wave function $\Phi_{\lambda}$ becomes orthogonal to the configurations $\{\Phi_c\}$.
\red{It is noted that each configuration $\Phi_{\lambda,n}$ is not imposed to be orthogonal to $\{\Phi_c\}$.
  With a small value of $\lambda$, $\Phi_{\lambda}$ is not fully orthogonal to $\{\Phi_c\}$,
  but it can be excited from the ground state with a small excitation energy.
  Hence, the configurations $\{\Phi_{\lambda,n}\}$ can contribute to the low-lying nuclear states
  and we adopt $\{\Phi_{\lambda,n}\}$ obtained with various values of $\lambda$.
  Finally, we superpose these configurations with angular-momentum projection in the wave function in Eq.~(\ref{eq:linear}).}

%%%%%%%%%%%%%%%%%%%%%%%%%%%%%%%%%%%%%%%%%%%%
We explain the calculation procedure as follows:
\begin{enumerate}
\itemsep=2mm
\item[(a)]
  We prepare the multiple AMD basis states with index $n=1,\cdots,N_{\rm b}$.
  Solving the multicool equation in Eq. (\ref{eq:multi_eq}), we obtain the ground-state configurations $\{\Phi_{n}\}$
  and define $\{\Phi_{c}\}$ with $c=1,\cdots,3N_{\rm b}$ using the rotations of $\{\Phi_n\}$.
  
\item[(b)]
  Introducing the pseudopotential $V_\lambda$ in the Hamiltonian with various values of $\lambda$ in Eq.~(\ref{eq:PSE}),
  we solve the multicool equation with the basis number $N_{\rm b}$
  and obtain the excited-state configurations $\{\Phi_{\lambda,n}\}$.

\item[(c)]
  We superpose $\{\Phi_n\}$ and $\{\Phi_{\lambda,n}\}$ with the angular-momentum and parity projections, and
  solve the generalized eigenvalue problem of the Hamiltonian matrix in Eq. (\ref{eq:eigen}). 
  \red{From this equation, we can determine the weights of the configurations in the individual eigenstates.
  These configurations include the mean-field and cluster structures.}
\end{enumerate}

\red{
The basis number $N_{\rm b}$ is determined to get the relevant configurations of the system as a convergence.
In the $p$-shell nuclei, $N_{\rm b}$ is typically around 15--20 in Eqs.~(\ref{eq:multi}) and (\ref{eq:multi2}) \cite{myo23b,myo25a},
and we can obtain the configurations of the mean-field and cluster states.}
The total basis number in the above step (c) is at most about 600 for $^{10}$C and $^{10}$Be before the $K$-mixing.

Some of the specific properties of the multicool framework are explained in Refs. \cite{myo23b,myo25a},
such as the resulting AMD configurations after the variation.
In the results of the variation, we often obtain the cluster configurations with the same cluster constituents, but with different intercluster distances.
This is considered to be the optimization of the relative wave function between clusters through the superposition of the configurations,
which is automatically performed in the multicool calculation.
Additionally, the pseudopotential $V_\lambda$ effectively generates the multiple configurations necessary for the excited states.
This is also a prominent point of the present method.

\subsection{Hamiltonian}\label{sec:ham}
In the Hamiltonian, we use the effective nuclear interactions,
which consist of the two-body central, spin-orbit, and Coulomb forces.
In this study, we use the Volkov No.2 for the central force \cite{volkov65} and the G3RS for the spin-orbit force \cite{tamagaki68,yamaguchi79}
according to the previous studies of $^{10}$Be \cite{itagaki00,itagaki02b,suhara10,shikata20,myo23b}.
We use the Majorana parameter $M$=0.6 and the Bartlett and Heisenberg parameters $B=H=0.125$ and 1600 MeV of the spin-orbit strength.
This Hamiltonian is the same as that used in other calculations \cite{itagaki00,itagaki02b,suhara10,shikata20}
and reproduces the systematic binding energies well of $^6$He, Li isotopes, and $^{9,10}$Be in the multicool calculations \cite{myo23b,myo25a}.
\red{
It is noted that 
in Ref. \cite{itagaki02b}, the authors discuss the dependence of the energy spectrum and $B(E2)$ of $^{10}$Be on the spin-orbit strength.
For other interactions, the Gogny D1S force has been employed for $^{10}$Be with AMD \cite{isaka15},
and the resulting energy spectrum is similar to that with the present interaction in our previous work \cite{myo23b}.}

Following the previous works \cite{itagaki00,itagaki02b,suhara10,shikata20,myo23b}, we set $\nu=0.235$ fm$^{-2}$ in Eq.~(\ref{eq:AMD}).
The total energy of the $\alpha$ particle is $-27.59$ MeV with the $(0s)^4$ configuration,
and the $\alpha+\alpha+p+p$ threshold energy of $^{10}$C is $-55.17$ MeV, which is close to the experimental value of $-56.59$ MeV.

In Table \ref{tab:ene_sub}, we list the total energies of the subsystems of $^{10}$C and $^{10}$Be obtained in the multicool calculation \cite{myo23b},
in comparison with the experimental values.
For unbound nuclei of $^6$Be and $^9$B, the bound-state approximation is adopted.
We also evaluate the energy differences between the mirror nuclei.
It is confirmed that the present Hamiltonian reproduces the experimental energies of $A=6$ and $A=9$ nuclei,
which are the threshold constituents of $A=10$ nuclei.

%%%%%%%%%%%%%%%%%%%%%%%%%%%%%% 
\begin{table}[t]
  \caption{
    Total energies of the ground states of $\alpha$, $^6$He, $^6$Be, $^9$Be, and $^9$B,
    and the energy differences $\Delta E$ between mirror nuclei. Units are in MeV.
  }\vspace*{0.1cm}
\label{tab:ene_sub}
\renewcommand{\arraystretch}{1.5}
\begin{tabular}{lrrrrrrrr}
\noalign{\hrule height 0.5pt}
                 &  Expt    && Present \\
\noalign{\hrule height 0.5pt}
$\alpha$                     & $-28.30$ && $-27.59$ \\ 
$^6$He ($0^+$)               & $-29.27$ && $-29.20$ \\ 
$^6$Be ($0^+$)               & $-26.93$ && $-26.87$ \\ 
$^9$Be ($3/2^-$)             & $-58.17$ && $-57.45$ \\ 
$^9$B  ($3/2^-$)             & $-56.32$ && $-55.61$ \\ 
\noalign{\hrule height 0.5pt}
$\Delta E$($^6$He --$^6$Be)  & $  2.35$ && $  2.33$ \\ 
$\Delta E$($^9$Be --$^9$B)   & $  1.85$ && $  1.85$ \\ 
\noalign{\hrule height 0.5pt}
\end{tabular}
\end{table}
%%%%%%%%%%%%%%%%%%%%%%%%%%%%%%

%%%%%%%%%%%%%%%%%%%%%%%%%%%%%% 
\begin{table}[t]
  \caption{
    Total energies and matter radii of the intrinsic state of $^{10}$C with positive parity
    in the multicool calculation in comparison with the single AMD basis calculation.
  }\vspace*{0.1cm}
\label{tab:multi}
\renewcommand{\arraystretch}{1.5}
\begin{tabular}{lrrrrrr}
\noalign{\hrule height 0.5pt}
                              &&   Single    && Multicool  \\
\noalign{\hrule height 0.5pt}
Energy (MeV)                  &&  $-44.73$   &&  $-50.55$  \\  
Radius (fm)                   &&     2.10    &&     2.21   \\
\noalign{\hrule height 0.5pt}
\end{tabular}
\end{table}
%%%%%%%%%%%%%%%%%%%%%%%%%%%%%%

%%%%%%%%%%%%%%%%%%%%%%%%%%%%%%%%%%%%%%%%%%%%
\section{Results}\label{sec:result}

\subsection{Energy variation} 

We perform the multicool calculation for $^{10}$C using $N_{\rm b}=16$ intrinsic AMD basis states in Eq.~(\ref{eq:multi}). 
The results of the energy variation in the intrinsic frame for the positive parity states are summarized in Table \ref{tab:multi}.
We compare the results with those from a single AMD calculation.
The energy gain due to the multiple bases is 5.8 MeV and the radius becomes large by 0.11 fm,
indicating the mixing of the bases with large radii.

\begin{figure}[t]
\centering
\includegraphics[width=7.5cm,bb=0 0 360 252]{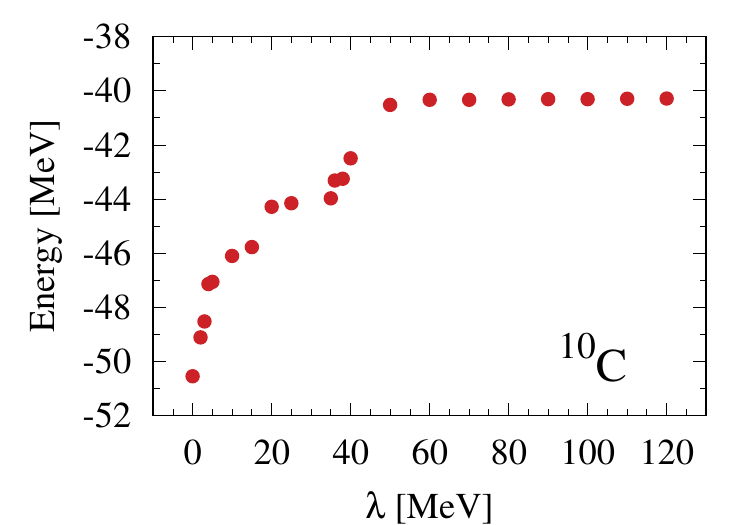}\\
\includegraphics[width=7.5cm,bb=0 0 360 252]{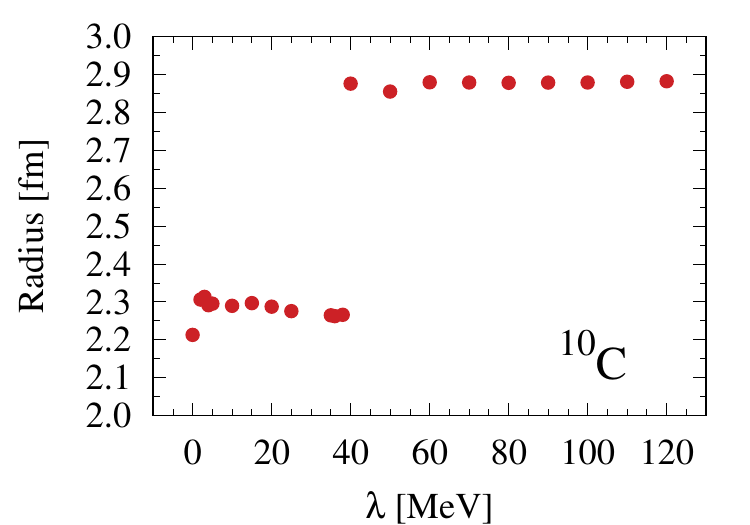}
\caption{
  Intrinsic energy (upper) and radius (lower) of $^{10}$C for positive parity state with the basis number $N_{\rm b}=16$ in the multicool calculation.
  The strength $\lambda$ of the pseudopotential changes.}
\label{fig:ene_positive}
\end{figure}

Next, we prepare the basis states for the excited states of $^{10}$C with the pseudopotential $V_\lambda$ keeping the basis number $N_{\rm b}=16$
using Eqs.~(\ref{eq:PSE}) and (\ref{eq:multi2}).
In Fig.~\ref{fig:ene_positive}, we show the results of the total energy and matter radius for the positive parity state of $^{10}$C
by changing the strength $\lambda$ in the pseudopotential in Eq.~(\ref{eq:PSE}).
At $\lambda=0$, the calculation corresponds to the state, shown in Table \ref{tab:multi}.
It is noted that the total energies are plotted omitting the contribution of the pseudopotential.

By increasing $\lambda$, the energy gradually increases and becomes stable at around $-40$ MeV beyond $\lambda=40$ MeV,
which indicates the excited states with the excitation energy of about 10 MeV.
The radius also increases to around 2.9 fm, which is larger than the ground-state radius by about 0.7 fm.
This indicates that the system is expanded spatially.

%%%%%%%%%%%%%%%%%%%%%%%%%%%%%%%%%%%%
%%%%%%%%%%%%%%%%%%%%%%%%%%%%%%%%%%%%

%%%%%%%%%%%%%%%%%%%%%%%%%%%%%%%%%%%%
\begin{figure}[t]
  \centering
\includegraphics[width=8.5cm,bb=0 0 249 163]{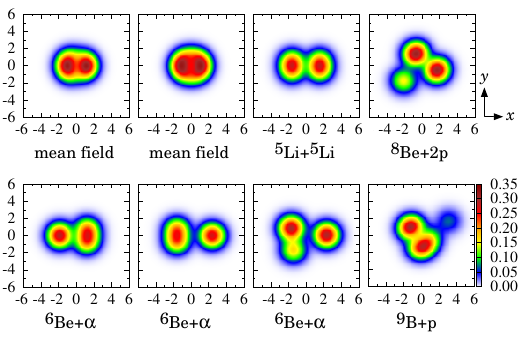} 
\\[-0.3cm]
\caption{
  Intrinsic density distributions of the representative configurations of $^{10}$C for the ground state. 
  Units of densities and axes are in fm$^{-3}$ and in fm, respectively.
  The lower label in each panel explains the configuration.
}
\label{fig:density1}
\end{figure}
%%%%%%%%%%%%%%%%%%%%%%%%%%%%%%%%%%%%
\begin{figure}[th]
  \centering
\includegraphics[width=8.5cm,bb=0 0 249 166]{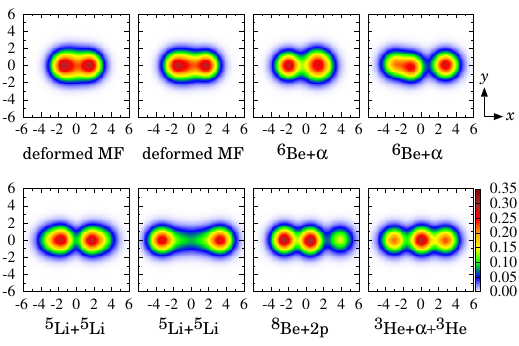} 
\\[-0.3cm]
\caption{
  Intrinsic density distributions of the representative configurations of $^{10}$C for the excited states
  with the pseudopotential. Units of densities and axes are in fm$^{-3}$ and in fm, respectively.
  The lower label in each panel explains the configuration.
}
\label{fig:density2}
\end{figure}

In Fig.~\ref{fig:density1}, we show the intrinsic density distributions of the representative eight configurations of
the ground state of $^{10}$C with the positive parity, shown in Table \ref{tab:multi}.
\red{The volume integral of each density distribution corresponds to the mass number.}
In the figure, we adjust the direction of the longest distribution to be the horizontal $x$ axis. 
One can confirm various density distributions of the mean-field structure, which is compact spatially,
and the different clusterings such as $^5$Li+$^5$Li, $^6$Be+$\alpha$, $^8$Be+$2p$, and  $^9$B+$p$. 
It is noted that the distance between clusters can be different, such as in the $^6$Be+$\alpha$ configurations.
This property corresponds to the superposition of the configurations with various cluster distances and indicates the optimization of the intercluster wave function of $^{10}$C.
In the present calculation,
the emergence of various cluster configurations is related to the presence of the threshold energies of the corresponding cluster emissions,
as will be shown later in the energy spectrum.
This property can be regarded as the threshold rule in the theoretical calculation \cite{ikeda68,myo23b,myo25a},
although the existence of the clustering states is not concluded experimentally in $^{10}$C.

In Fig.~\ref{fig:density2}, we also show the intrinsic density distributions of the representative eight configurations of
the excited states of $^{10}$C, which are obtained in the stable region with respect to the strength $\lambda$ of the pseudopotential,
as shown in Fig. \ref{fig:ene_positive}.
The configurations commonly show the elongated linear-chain shapes, which are the same results in the excited states of $^{10}$Be \cite{myo23b}.
In addition to the cluster states involving two $\alpha$ particles, the deformed mean-field (MF) state and the $^3$He+$\alpha$+$^3$He state are obtained.
These configurations are largely mixed in the $0^+_2$ states and also its band members in $^{10}$C, as is discussed later.

%%%%%%%%%%%%%%%%%%%%%%%%%%%%
\subsection{Energy spectrum}

%%%%%%%%%%%%%%%%%%%%%%%%%%%%%% 
\begin{table}[t]
  \caption{
    Ground-state properties of $^{10}$C ($0^+_1$) and $^{10}$Be ($0^+_1$) ; the total energies in units of MeV and
    the radii of matter ($r_{\rm m}$), proton ($r_{\rm p}$), neutron ($r_{\rm n}$), and charge ($r_{\rm ch}$)  in units of fm,
    in comparison with the experimental values \cite{ozawa96,ozawa01,tanihata88,nortershauser09} and the four-body cluster model (4CM) \cite{ogawa00}.
  }\vspace*{0.1cm}
\label{tab:ground}
\renewcommand{\arraystretch}{1.5}
\begin{tabular}{lcrrrr}
\noalign{\hrule height 0.5pt}
Energy              &&  Expt       && Present     \\ 
\noalign{\hrule height 0.5pt}                                   
$^{10}$C            &&  $-60.32$   && $-59.32$    \\
$^{10}$Be           &&  $-64.98$   && $-63.75$    \\
$\Delta E$          &&  $  4.66$   && 4.43        \\
\noalign{\hrule height 0.5pt}
\end{tabular}
\\[0.2cm]
\begin{tabular}{lcrrrrrrr}
\noalign{\hrule height 0.5pt}
           &             &&  Expt       && Present     && 4CM      \\ 
\noalign{\hrule height 0.5pt}                                               
$^{10}$C   & $r_{\rm m}$ &&  2.27(3)    && 2.36        && 2.33     \\
           & $r_{\rm p}$ &&  2.31(3)    && 2.44        && 2.41     \\
           & $r_{\rm n}$ &&  2.22(3)    && 2.23        && 2.20     \\
           & $r_{\rm ch}$&&  2.44(3)    && 2.56        && 2.54     \\
\noalign{\hrule height 0.5pt}                                               
$^{10}$Be  & $r_{\rm m}$ &&  2.30(2)    && 2.33        && 2.28     \\
           & $r_{\rm p}$ &&  2.24(2)    && 2.21        && 2.17     \\
           & $r_{\rm n}$ &&  2.34(2)    && 2.40        && 2.35     \\
           & $r_{\rm ch}$&&  2.357(18)  && 2.35        && 2.31     \\
\noalign{\hrule height 0.5pt}
\noalign{\hrule height 0.5pt}
\end{tabular}
\end{table}
%%%%%%%%%%%%%%%%%%%%%%%%%%%%%%

\red{
We superpose the AMD configurations with the angular-momentum and parity projections,
where the configurations are obtained in the variations with the multiple bases.
We include the AMD configurations with the various values of $\lambda$ in the pseudopotential, as shown in Fig.~\ref{fig:ene_positive}.}
Solving the eigenvalue problem of the Hamiltonian matrix,
we obtain the ground $0^+_1$ state of $^{10}$C, the properties of which are summarized in Table \ref{tab:ground}.
The total energy and the radii of the ground state of $^{10}$C ($0^+_1$) are consistent with the experimental values.
We also compare the results with those of $^{10}$Be ; The energy difference $\Delta E$ between $^{10}$C and $^{10}$Be is close to the experimental value.
In our calculation, the radii of $^{10}$C are larger than those of $^{10}$Be due to the Coulomb repulsion.
Our charge radius of $^{10}$Be reproduces the experimental value, which is observed precisely in the isotope-shift measurements \cite{nortershauser09}.
The radial property of the two nuclei almost agrees with the $\alpha+\alpha+N+N$ four-body cluster model \cite{ogawa00}.
Our results give slightly larger values of the radii of the two nuclei than those in Ref. \cite{ogawa00}.

%%%%%%%%%%%%%%%%%%%%%%%%%%%%%%%%%%%%
\begin{figure*}[tbh]
\centering

\includegraphics[width=17.7cm,bb=0 0 685 252]{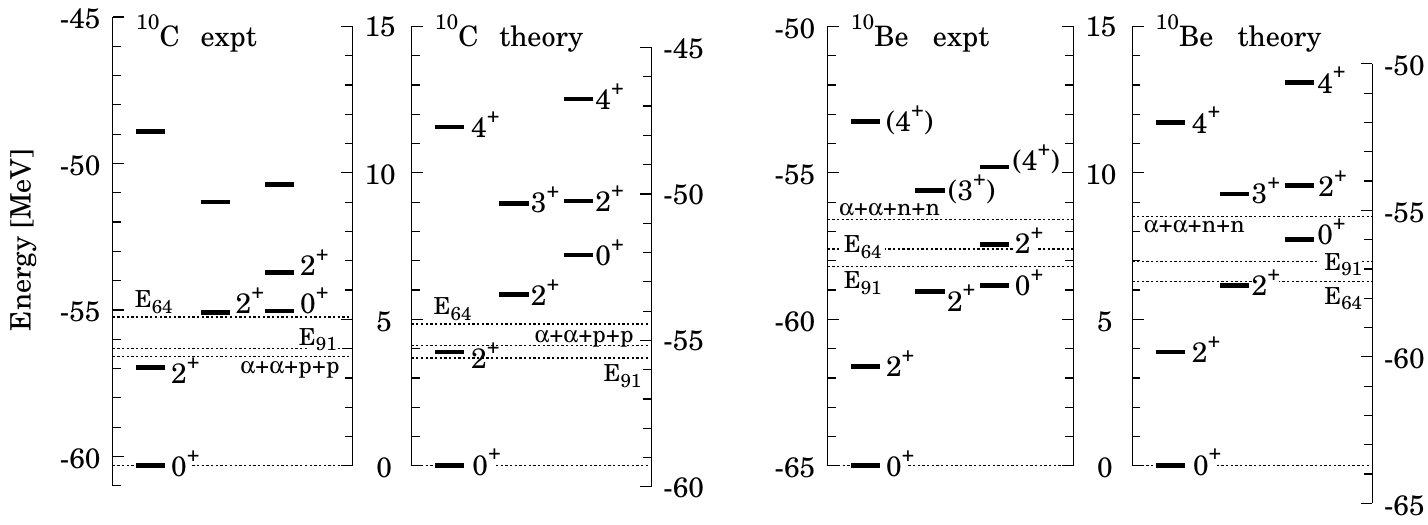}
\caption{
  Energy spectra of $^{10}$C (left two panels) and $^{10}$Be (right two panels) in the experiments \cite{tilley04,sobotka24}
  and the multicool calculation in units of MeV.
  In each spectrum, the left-hand and right-hand sides of the measures represent the total energies and the measures
  in the middle represent the excitation energy.
  The threshold energies are shown with the dotted horizontal lines
  for $\alpha$+$\alpha$+$N$+$N$, $^6$He/$^6$Be+$\alpha$ denoted as $E_{64}$, and $^9$Be+$n$/$^9$B+$p$ denoted as $E_{91}$.}
\label{fig:ene_GCM}
\end{figure*}
%%%%%%%%%%%%%%%%%%%%%%%%%%%%%%%%%%%%

We show the excitation energy spectra of $^{10}$C and $^{10}$Be in Fig. \ref{fig:ene_GCM}.
\red{Our results reproduce the experimental levels well, although the spin assignment is not fully done in the experiment.}
The $2^+_2$ is dominantly the $K^\pi=2^+$ state and the $0^+_2$ state is dominantly the linear-chain state; the representative configurations are
shown in Fig. \ref{fig:density2}.
The excitation energy of the $0^+_2$ state is 7.22 MeV and is larger than the experimental value of 5.38 MeV by about 1.8 MeV in $^{10}$C.
This is the same trend as shown in $^{10}$Be; the theoretical excitation energy is 7.72 MeV,
which is larger than the experimental value of 6.18 MeV by about 1.5 MeV.
If we weaken the strength of the spin-orbit force, the excitation energy of the $0^+_2$ state becomes lower
because of the reduction of the attraction of the spin-orbit force in the ground $0^+_1$ state coming from the $p_{3/2}$ configuration.
It is noted that the excitation energy of the $4^+_1$ state is lower than that of $4^+_2$ in the calculations of $^{10}$C and $^{10}$Be,
but this order might be changed in the experiments, although the spin assignment is not settled yet in the experiment.

In Table \ref{tab:quanta},
we list the properties of the eight states of $^{10}$C and $^{10}$Be; expectation values of the principal quantum number operator $N$
of the harmonic oscillator, where the operator $N$ is defined in terms of the operators of kinetic energy and radius
and $\hbar \omega=2\nu \hbar^2/m$ with a nucleon mass $m$:
\begin{equation}
  \begin{split}
  N\coloneqq \sum_{i=1}^A \left(\dfrac{\bm{p}_i^2}{4\hbar^2\nu}+\nu \bm{r}^2_i - \dfrac32 \right).
  \end{split}
  \label{eq:quanta}
\end{equation}
The values of $\bra N \ket$ are useful for estimating the amounts of excitation of the states in the picture of the harmonic-oscillator shell model.
We also show three kinds of radii of the states to know the spatial extension.

In the upper table of Table \ref{tab:quanta}, we show the results of $^{10}$C.
In the ground $0^+_1$ state, the value of $\bra N\ket$ is 7.51, which is larger than the lowest quanta of 6 with the $p$ shell configuration.
For the matter radius, the calculation is 2.36 fm, which is larger than the value of $\sqrt{39/(40\nu)}=2.04$ fm with the lowest quanta of 6, by 0.32 fm.
These results indicate inclusion of the spatial correlations of the deformation or clustering beyond the $p$ shell
in the ground state as shown in the densities in Fig. \ref{fig:density1}.
In the table, it is found that three states of $0^+_2$, $2^+_3$, and $4^+_2$ in $^{10}$C commonly show the large value of $\bra N \ket $ with around 13--14
and also the large matter radii of around 2.9 fm. The proton and neutron radii are also larger than those of other states.
These results come from the elongated linear-chain structure as illustrated in the density distributions shown in Fig. \ref{fig:density2}.

In the lower table of Table \ref{tab:quanta}, we list the properties of $^{10}$Be.
The values are very similar to those of $^{10}$C with the exchange of proton and neutron,
which suggests a good isospin symmetry in the two nuclei for quanta and radius.

We calculate the expectation values of the squared spin operator $\bm{S}^2$ for the states of $^{10}$C and $^{10}$Be,
where $\bm{S}=\sum_{i=1}^A \bm{s}_i$ with nucleon spin operator $\bm{s}_i$.
We also show the values of the proton and neutron components as $\bm{S}_p^2$ and $\bm{S}_n^2$, respectively, where $\bm{S}=\bm{S}_p+\bm{S}_n$.
The values of $\bra \bm{S}^2 \ket$ are useful for discussing the spin component of the state.
If $\alpha$ cluster is formed with an $s$-wave configuration and spin saturation, its contribution to $\bra \bm{S}^2 \ket$ becomes zero.

\red{
In Table \ref{tab:spin_square}, we show the results of $\bra \bm{S}^2 \ket$ for each state of the two nuclei.
It is found that in $^{10}$C, spin contribution mainly comes from the proton part and the neutron contribution is small for every state.}
The small value of $\bra \bm{S}_n^2\ket$ can be related to the $\alpha$ cluster formation in $^{10}$C.
This tendency becomes significant in the $0^+_2$, $2^+_3$, and $4^+_2$ states with the value of 0.02, having commonly the linear-chain structure,
as shown in Fig. \ref{fig:density2}.

In $^{10}$Be, the same properties of the squared spin values can be confirmed in the exchange of the proton and neutron parts of $^{10}$C.
This also indicates a good isospin symmetry in the two nuclei, together with the quanta and radii in Table \ref{tab:quanta}.

%%%%%%%%%%%%%%%%%%%%%%%%%%%%%% 
\begin{table}[t]
\centering
\caption{
  Expectation values of the principal quantum number operator $N$ of $^{10}$C (upper table) and $^{10}$Be (lower table)
  in the multicool calculation.
  We also show the radii of matter ($r_{\rm m}$), proton ($r_{\rm p}$), and neutron ($r_{\rm n}$) in units of fm. 
  }\vspace*{0.1cm}
\label{tab:quanta}
\renewcommand{\arraystretch}{1.5}
\begin{tabular}{crrrrrrrrrrrrrrrrrrrr}
\noalign{\hrule height 0.5pt}
$^{10}$C     && $0^+_1$ && $0^+_2$ && $2^+_1$ && $2^+_2$ && $2^+_3$  && $3^+$ && $4^+_1$ && $4^+_2$  \\
\noalign{\hrule height 0.5pt}                                                            
$\bra N\ket$ && 7.51    && 13.32   &&  7.55   && 7.66    && 13.56    && 7.75  && 8.28    &&  13.52   \\
$r_{\rm m}$  && 2.36    &&  2.90   &&  2.36   && 2.40    &&  2.92    && 2.40  && 2.44    &&   2.91   \\
$r_{\rm p}$  && 2.44    &&  3.02   &&  2.45   && 2.47    &&  3.04    && 2.48  && 2.53    &&   3.03   \\
$r_{\rm n}$  && 2.23    &&  2.70   &&  2.23   && 2.29    &&  2.73    && 2.28  && 2.30    &&   2.73   \\
\noalign{\hrule height 0.5pt}
\end{tabular}
\\[0.5cm]
%%%%%%%%%%%%%%%%%%%%%%%%%%%%%%
\begin{tabular}{crrrrrrrrrrrrrrrrrrrr}
\noalign{\hrule height 0.5pt}
$^{10}$Be    && $0^+_1$ && $0^+_2$ && $2^+_1$ && $2^+_2$ && $2^+_3$  && $3^+$  && $4^+_1$ && $4^+_2$  \\
\noalign{\hrule height 0.5pt}                                                             
$\bra N\ket$ && 7.35    && 13.10   &&  7.37   && 7.82    && 12.37    && 7.82   && 7.83    &&  13.36   \\
$r_{\rm m}$  && 2.33    &&  2.88   &&  2.33   && 2.42    &&  2.81    && 2.41   && 2.39    &&   2.90   \\
$r_{\rm p}$  && 2.21    &&  2.70   &&  2.22   && 2.30    &&  2.66    && 2.28   && 2.26    &&   2.73   \\
$r_{\rm n}$  && 2.40    &&  2.99   &&  2.40   && 2.50    &&  2.91    && 2.49   && 2.47    &&   3.00   \\
\noalign{\hrule height 0.5pt}
\end{tabular}
\end{table}
%%%%%%%%%%%%%%%%%%%%%%%%%%%%%%
\begin{table}[t]
\centering
\caption{
  Expectation values of the squared spin operator $\bm{S}^2$, where $\bm{S}$ is a sum of the nucleon spin operator.
  We also show the proton $\bm{S}_p^2$ and neutron $\bm{S}_n^2$ parts.
  }\vspace*{0.1cm}
\label{tab:spin_square}
\renewcommand{\arraystretch}{1.5}
\begin{tabular}{lcrrrrrrrrrrrrrrrrrrrr}
\noalign{\hrule height 0.5pt}
$^{10}$C               && $0^+_1$ && $0^+_2$ && $2^+_1$ && $2^+_2$ && $2^+_3$  && $3^+$  && $4^+_1$ && $4^+_2$  \\
\noalign{\hrule height 0.5pt}                                                            
$\bra \bm{S}^2 \ket$   && 0.62    &&  0.21   &&  0.62   && 0.32    &&  0.21    && 0.27   && 0.44    &&  0.25   \\
$\bra \bm{S}_p^2 \ket$ && 0.54    &&  0.19   &&  0.53   && 0.27    &&  0.19    && 0.23   && 0.33    &&  0.22   \\
$\bra \bm{S}_n^2 \ket$ && 0.08    &&  0.02   &&  0.09   && 0.04    &&  0.02    && 0.04   && 0.10    &&  0.02   \\
\noalign{\hrule height 0.5pt}
\end{tabular}
\\[0.5cm]
%%%%%%%%%%%%%%%%%%%%%%%%%%%%%%
\begin{tabular}{lcrrrrrrrrrrrrrrrrrrrr}
\noalign{\hrule height 0.5pt}
$^{10}$Be              && $0^+_1$ && $0^+_2$ && $2^+_1$ && $2^+_2$ && $2^+_3$  && $3^+$  && $4^+_1$ && $4^+_2$  \\
\noalign{\hrule height 0.5pt}                                                             
$\bra \bm{S}^2 \ket$   && 0.64    &&  0.22   &&  0.65   && 0.18    &&  0.27    && 0.15   && 0.50    &&  0.23   \\
$\bra \bm{S}_p^2 \ket$ && 0.09    &&  0.02   &&  0.10   && 0.05    &&  0.02    && 0.05   && 0.13    &&  0.02   \\
$\bra \bm{S}_n^2 \ket$ && 0.55    &&  0.20   &&  0.55   && 0.13    &&  0.25    && 0.11   && 0.36    &&  0.21   \\
\noalign{\hrule height 0.5pt}
\end{tabular}
\end{table}
%%%%%%%%%%%%%%%%%%%%%%%%%%%%%%
%
\subsection{Monopole transition}

From the intrinsic density distribution of the configurations of $^{10}$C, the cluster states are confirmed
in Figs. \ref{fig:density1} and \ref{fig:density2}.
It has been discussed that cluster states tend to give a large monopole transition strength,
which can be a signature to prove the clustering in the nuclear states \cite{yamada08}.
According to this discussion, we calculate the monopole transition strengths of $^{10}$C and $^{10}$Be, which are summarized in Table \ref{tab:mono},
where the transition operator is defined as $\sum_{i=1}^A (\bm{r}_i-\bm{r}_{\rm G})^2$ using the center-of-mass coordinate $\bm{r}_{\rm G}$.

In the table, the transitions of the isoscalar (IS0), proton part corresponding to the electric transition ($E0$), and neutron part are shown.
For comparison, we estimate the single-particle strength of $\sqrt{5/(8\nu^2)}=3.36$ fm$^2$ from the $0p$ to $1p$ orbits \cite{ito14}.
For IS0, three kinds of transitions; $0^+_1 \to 0^+_2$, $2^+_1 \to 2^+_3$, and $4^+_1 \to 4^+_2$, show large values
and are comparable to the single-particle strength.
The results support the spatial extension due to the elongated linear-chain structure in $^{10}$Be and $^{10}$C with large quanta.
For the isospin symmetry in the two nuclei, the $0^+$ transition shows a good symmetry in the exchange of proton and neutron parts.
The $2^+$ transitions show a good symmetry with a slight difference.
For the $4^+$ state, $^{10}$C shows the larger values than those of $^{10}$Be
in each mirror component and their ratios are almost the same at around 1.4.

%%%%%%%%%%%%%%%%%%%%%%%%%%%%%% 
\begin{table}[t]
  \caption{
    Magnitudes of the matrix elements $|M|$ of the monopole transitions of $^{10}$Be and $^{10}$C in units of fm$^2$,
    for the isoscalar part (IS0), proton part corresponding to the electric transition ($E0$), and neutron part,
    where $M(\mbox{IS0})=M({\rm proton})+M({\rm neutron})$.
  }\vspace*{0.1cm}
\label{tab:mono}
\renewcommand{\arraystretch}{1.5}
\begin{tabular}{lccccccccc}
\noalign{\hrule height 0.5pt}
                  && \multicolumn{3}{c}{$^{10}$Be} & & \multicolumn{3}{c}{$^{10}$C} \\ \cline{3-5} \cline{7-9}
                  && IS0      & Proton & Neutron   &~& IS0      & Proton & Neutron  \\
\noalign{\hrule height 0.5pt}
$0^+_1 \to 0^+_2$ && 3.23     & 1.12   & 2.11      & & 3.22     & 2.11   & 1.11    \\
$2^+_1 \to 2^+_2$ && 1.17     & 0.36   & 0.81      & & 1.10     & 0.77   & 0.33    \\
$2^+_1 \to 2^+_3$ && 2.38     & 0.84   & 1.55      & & 2.70     & 1.76   & 0.94    \\
$2^+_2 \to 2^+_3$ && 0.10     & 0.03   & 0.07      & & 0.16     & 0.09   & 0.07    \\
$4^+_1 \to 4^+_2$ && 5.64     & 1.99   & 3.65      & & 7.83     & 5.10   & 2.73    \\
\noalign{\hrule height 0.5pt}
\end{tabular}
\end{table}
%%%%%%%%%%%%%%%%%%%%%%%%%%%%%%

\subsection{Quadrupole moment}

%%%%%%%%%%%%%%%%%%%%%%%%%%%%%% 
\begin{table}[t]
  \caption{
    Quadrupole moments of the $2^+$ and $4^+$ states of $^{10}$Be and $^{10}$C for proton and neutron parts
    in comparison with the calculations of AMD \cite{kanada97}, MCSM \cite{liu12}, and GFMC \cite{mccutchan12}.
    Units are in fm$^2$.
  }\vspace*{0.1cm}
\label{tab:Q}
\renewcommand{\arraystretch}{1.5}
\begin{tabular}{llrrrrrrrrrrr}
\noalign{\hrule height 0.5pt}  \vspace*{-0.10cm}
        & && \multicolumn{2}{c}{$^{10}$Be} && \multicolumn{2}{c}{$^{10}$C} \\ \cline{4-5} \cline{7-8}
        &         &&  Proton   & Neutron   &&   Proton & Neutron  \\
\noalign{\hrule height 0.5pt}                                                                   
$2^+_1$~& Present && $-5.13$   & $-2.62$   && $-2.21$  &  $-4.97$ \\
        & AMD     && $-6.5$    &           && $-3.8$   &          \\
        & MCSM    && $-5.9$    & $-3.1$    &&          &          \\  
        & GFMC    && $-6.7(1)$ &           && $-2.7(2)$&          \\  
\noalign{\hrule height 0.5pt}                                                                   
$2^+_2$~& Present && $ 5.45$   &  $ 3.68$  && $ 2.56$  &  $ 3.81$ \\
        & MCSM    && $ 5.8$    &  $ 3.6$   &&          &          \\  
        & GFMC    && $ 4.5(1)$ &           && $-0.9(3)$&          \\  
\noalign{\hrule height 0.5pt}                                                                   
$2^+_3$ & Present && $-9.24$   & $-18.00$  && $-21.74$ & $-10.79$ \\
$4^+_1$ & Present && $-5.22$   & $  1.12$  && $  0.41$ & $ -4.79$ \\
$4^+_2$ & Present && $-13.10$  & $-25.15$  && $-25.59$ & $-13.39$ \\
\noalign{\hrule height 0.5pt}
\end{tabular}
\end{table}
%%%%%%%%%%%%%%%%%%%%%%%%%%%%%%

We discuss the deformation properties of $^{10}$C and $^{10}$Be in terms of their quadrupole moments.
\red{This analysis is useful to understand the anomaly of the electric quadrupole transitions in $^{10}$C and $^{10}$Be.}
In Table \ref{tab:Q}, we show the quadrupole moments of the $2^+$ and $4^+$ states of $^{10}$Be and $^{10}$C.
In addition to the proton part as an electric quadrupole moment, we also calculate the neutron part to discuss the isospin symmetry.
There has been no experimental data in the two nuclei and we compare our results with other theoretical calculations of AMD \cite{kanada97},
Monte Carlo shell-model (MCSM) \cite{liu12}, and Green's function Monte Carlo (GFMC) \cite{mccutchan12}.
Our results are similar to those of MCSM for the $2^+_{1,2}$ states of $^{10}$Be.
In MCSM, the Coulomb force is not included and then $^{10}$Be and $^{10}$C give the same numbers in the exchange of the proton and neutron parts.
Recently, there is an attempt of the experimental analysis of the quadrupole moments of the $2^+_1$ states in the two nuclei
with the help of the no-core shell model \cite{li24}.

The $2^+_1$ states, which have dominantly $K^\pi=0^+$, give the negative values of proton and neutron parts in the two nuclei.
In $^{10}$Be, the proton part is larger than the neutron part, indicating the spherical property of the neutrons in $^{10}$Be.
This is considered to come from the sub-closed nature of the $p_{3/2}$ orbit of neutrons in $^{10}$Be.
The same tendency is confirmed in the protons of $^{10}$C.
It is found that the proton quadrupole moment gives a smaller magnitude than that of neutron in $^{10}$Be by about 0.4 fm$^2$.
\red{
  This suggests a more spherical nature of protons in $^{10}$C than of neutrons in $^{10}$Be.
  This property of $^{10}$C becomes the origin of the small electric quadrupole transition of $^{10}$C from $2^+_1$ to $0^+_1$,
  and explains the anomaly of quadrupole transitions in $^{10}$C and $^{10}$Be, as is discussed in the next subsection.}

%%%%%%%%%%%%%%%%%%%%%%%%%%%%%%%%%%%%
\begin{figure}[t]
  \centering
\includegraphics[width=8.5cm,bb=0 0 249 94]{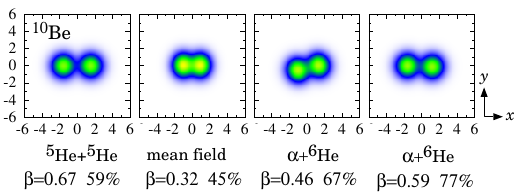}\\[0.10cm]
\includegraphics[width=8.5cm,bb=0 0 249 95]{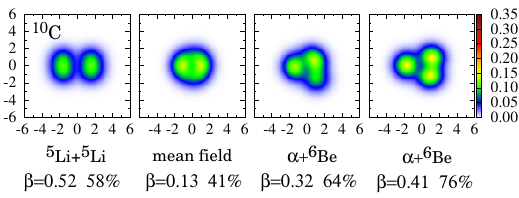}\\[-0.1cm]
\caption{
  Proton density distributions of the representative configurations of the $2^+_1$ states of $^{10}$Be (upper four panels) and $^{10}$C (lower four panels).
  Units of densities and axes are in fm$^{-3}$ and in fm, respectively. The lower label in each panel explains the configuration.
  The deformation parameters $\beta$ of protons and the squared overlaps with the $2^+_1$ state in units of \% are shown at the bottom of each panel.
}
\label{fig:density1p}
\end{figure}

To understand the deformation properties of the $2^+_1$ states of $^{10}$C and $^{10}$Be intuitively,
we compare the intrinsic proton density distributions of the representative AMD configurations in the $2^+_1$ states of $^{10}$Be and $^{10}$C in Fig. \ref{fig:density1p}.
In each panel, we show the quadrupole deformation parameter $\beta$ of protons, the definition of which is given in Ref. \cite{dote97},
and the squared overlaps of the configurations with respect to the total $2^+_1$ wave function in units of \%.
It is confirmed that four protons in $^{10}$Be form the 2p-2p structure in the horizontal $x$ direction and their relative distance depends on the configurations.
This structure originates from the $\alpha$-$\alpha$ clustering with neutrons and enhances the magnitude of the proton quadrupole moment in $^{10}$Be, as shown in Table \ref{tab:Q}.

However, the distributions of six protons of $^{10}$C look less deformed than those of $^{10}$Be,
This property consistently affects the values of $\beta$ of protons in the two nuclei; the values of $^{10}$C are smaller than those of $^{10}$Be between the corresponding mirror configurations with similar weights.
These results contribute to reduce the magnitude of the proton quadrupole moment in $^{10}$C.

The $2^+_2$ state, which has dominantly $K^\pi=2^+$, gives the positive values of proton and neutron parts in the two nuclei.
In these states, the proton part is larger than the neutron part in $^{10}$Be and the opposite results are obtained in $^{10}$C.
This indicates the large deformations in the protons (neutrons) in $^{10}$Be ($^{10}$C), similar to the results of the $2^+_1$ states.
It is also found that the values of $^{10}$C are smaller than those of $^{10}$Be in the exchange of proton and neutron.
This reduction indicates the small proton quadrupole deformation of the $2^+_2$ state in $^{10}$C.

For the $2^+_3$ state, this state involves the elongated linear-chain configuration in the two nuclei \cite{myo23b},
as shown in Fig. \ref{fig:density2} for $^{10}$C.
These configurations enhance the quadrupole moment with large negative numbers, as shown in Table \ref{tab:Q}. 
In this case, the values of $^{10}$C are larger than those of $^{10}$Be in magnitude in their mirror components.
This is related to the larger matter radius of $^{10}$C with 2.92 fm than that of $^{10}$Be with 2.81fm for the $2^+_3$ states,
as shown in Table \ref{tab:quanta}.
In fact, the proton/neutron radii are 3.04 fm / 2.73 fm for $^{10}$C and 2.66 fm / 2.91 fm for $^{10}$Be, respectively, in the calculation.

For the $4^+_1$ state, the proton part of $^{10}$Be is similar to the neutron part of $^{10}$C.
The neutron part of $^{10}$Be and the proton part of $^{10}$C show small positive values indicating small deformations
and this trend is the same as the results of the $2^+_1$ and $2^+_2$ states.
For the $4^+_2$ state, similar to the $2^+_3$ state, every component shows a large negative value
and this property is consistent with the linear-chain configurations in $^{10}$C and $^{10}$Be \cite{myo23b}.
The values are very similar to their mirror components in the two nuclei.
This indicates that the isospin symmetry is retained well.
This property can be seen in the radii of proton and neutron of the two nuclei in Table \ref{tab:quanta};
the proton/neutron radii are 3.03 fm / 2.73 fm for $^{10}$C and 2.73 fm / 3.00 fm for $^{10}$Be, respectively.

%%%%%%%%%%%%%%%%%%%%%%%%%%%%%% 
\begin{table*}[th]
  \caption{
    Electric quadrupole transition strength $B(E2)$ of $^{10}$Be and $^{10}$C in the multicool calculations
    in comparison with those of the experiments \cite{tilley04,mccutchan09,mccutchan12} and other calculations \cite{liu12,mccutchan12}
    including the molecular orbital model of $\alpha$+$\alpha$+$N$+$N$ (MO) \cite{itagaki00,itagaki02a}.
    Units are in $e^2 {\rm fm}^4$.
    In the last column, the ratios of the present results of $^{10}$C to those of $^{10}$Be are shown.
  }\vspace*{0.1cm}
\label{tab:E2}
\renewcommand{\arraystretch}{1.5}
\begin{tabular}{clllllllllllllrrr}
\noalign{\hrule height 0.5pt}
                 && \multicolumn{5}{c}{$^{10}$Be}                    && \multicolumn{5}{c}{$^{10}$C} && Ratio~ \\ \cline{3-7} \cline{9-13} \cline{15-15}
                 && Expt         &  Present & MO    & MCSM &  GFMC  &~~& Expt    & Present & MO    & MCSM  & GFMC     &~~& C/Be \\
\noalign{\hrule height 0.5pt}                                                                      
$2^+_1\to 0^+_1$ &&  $9.2(3)$    &  7.85    & 11.26 & 9.29 &  8.8(4) && 8.8(3)   & 8.84    & 12.8  & 9.30  & 15.3(1.4)&& 1.13  \\
$2^+_1\to 0^+_2$ &&  $0.66(24)$  &  0.09    &  0.23 &      &         &&          & 0.54    &       &       &          && 5.76  \\
\noalign{\hrule height 0.5pt}                                                                      
$2^+_2\to 0^+_1$ &&  $0.11(2)$   &  0.37    &  0.44 & 0.32 &  1.8(1) &&          & 1.21    &  1.3  & 2.15  &  0.2(1)  && 3.31  \\
$2^+_2\to 0^+_2$ &&              &  0.0005  &  0.00 &      &         &&          & 0.0013  &       &       &          && 2.65  \\
\noalign{\hrule height 0.5pt}                                                                     
$2^+_3\to 0^+_1$ &&              &  0.05    &  0.19 &      &         &&          & 0.40    &       &       &          && 8.79  \\
$2^+_3\to 0^+_2$ &&              &  23.1    & 35.56 &      &         &&          & 114.1   &       &       &          && 4.95  \\
\noalign{\hrule height 0.5pt}                                                                     
$2^+_2\to 2^+_1$ &&              &  2.56    &  3.99 & 3.28 &         &&          & 8.89    &  17   & 12.81 &          && 3.47  \\
$2^+_3\to 2^+_1$ &&              &  0.08    &       &      &         &&          & 0.71    &       &       &          && 8.53  \\
\noalign{\hrule height 0.5pt}
\end{tabular}
\end{table*}
%%%%%%%%%%%%%%%%%%%%%%%%%%%%%%
%
\subsection{Quadrupole transition}
We summarize the electric quadrupole transition strengths $B(E2)$ associated with the $2^+$ states of $^{10}$Be and $^{10}$C in Table \ref{tab:E2},
including the experimental data \cite{tilley04,mccutchan09} and other calculations \cite{itagaki00,itagaki02a,liu12,mccutchan12}.
In the overall trends, among the theories, our results are similar to those of the molecular orbital (MO) model and MCSM in the two nuclei.
In particular, the present model with the multiple AMD configurations is a similar framework to MO as a cluster model,
and then the description of the states such as the linear-chain structure is commonly obtained in the two methods.

\red{For the transitions of $2^+_1\to 0^+_1$ in the two nuclei, our results almost reproduce the experiments.}
In the comparison with other calculations, in $^{10}$C, there are sizable differences and our value is close to that of MCSM.
It is noted that there is a recent analysis of $B(E2)$ of the two nuclei in {\it ab initio} no-core configuration interaction calculation \cite{caprio25}.
For the transition of $2^+_2\to 0^+_1$, our calculations provide the smaller values in the two nuclei in comparison with those of $2^+_1\to 0^+_1$.
This tendency agrees with the experiment of $^{10}$Be, and for $^{10}$C our result can be a prediction.
Among the theories, our results are close to those of MO in both nuclei.
For the transition of $2^+_3\to 0^+_2$, our calculations provide the large values in the two nuclei.
One of the possible explanations of the results is the large radii of the initial and final states, as shown in Table \ref{tab:quanta},
which can enhance the quadrupole transition strengths, including the squared radius in the transition operator.
The value of $^{10}$C is larger than that of $^{10}$Be because of the large proton number and larger proton radii of $^{10}$C than those of $^{10}$Be
shown in Table \ref{tab:quanta}.

We discuss the isospin symmetry in $B(E2)$ of the two nuclei.
In the last column of Table \ref{tab:E2}, we show the ratios of $B(E2)$ of $^{10}$C to that of $^{10}$Be in each mirror state,
which are useful to understand the isospin symmetry of $B(E2)$ in the two nuclei.
It is found that the ratio of the transition of $2^+_1 \to 0^+_1$ shows 1.13, which is a different property from other transitions \red{as an anomaly}.
The transitions except for the $2^+_1\to 0^+_1$, show the enhancement of the values from $^{10}$Be to $^{10}$C.
This is naively reasonable due to the increase of the proton number from $^{10}$Be to $^{10}$C.
For the $2^+_1\to 0^+_1$ case, the ratio is close to unity, and this situation is commonly seen in MCSM \cite{liu12} and the experiments \cite{mccutchan09,mccutchan12}, but not in GFMC \cite{mccutchan12}.
In GFMC, the value of $^{10}$C is large, and it is discussed that the three-body force affects the magnitude of this transition \cite{mccutchan12}.
\red{
In the present calculation, we discuss that in the quadrupole moments of the $2^+_1$ state of $^{10}$C,
the proton part is smaller than the neutron part, as is shown in Table \ref{tab:Q}, 
which indicates smaller deformation of protons than that of neutrons in $^{10}$C.
This property results in the small $B(E2;2^+_1 \to 0^+_1)$ of $^{10}$C, and becomes the origin of the $B(E2)$ anomaly in $^{10}$C and $^{10}$Be.}

%%%%%%%%%%%%%%%%%%%%%%%%%%%%%% 
\begin{table}[t]
  \caption{
    Quadrupole transition strength of $^{10}$Be and $^{10}$C in the multicool calculation.
    Isoscalar (IS2), isovector (IV2), proton (P2), and neutron (N2). Units are in fm$^4$.
    Values of P2 are the same as listed in Table \ref{tab:E2}.
  }\vspace*{0.1cm}
  \label{tab:IS2}
\renewcommand{\arraystretch}{1.5}
\begin{tabular}{cllllllllllll}
\noalign{\hrule height 0.5pt}
                 && \multicolumn{4}{c}{$^{10}$Be}     && \multicolumn{4}{c}{$^{10}$C}  \\ \cline{3-6} \cline{8-11}
                 && IS2    & IV2    & P2   & N2       && IS2   & IV2   & P2    & N2    \\
\noalign{\hrule height 0.5pt}
$2^+_1\to 0^+_1$ && 31.40  & 0.00   & 7.85 & 7.85     && 33.54 & 0.025  & 8.84  & 7.94  \\
$2^+_2\to 0^+_1$ && 0.57   & 3.86   & 0.37 & 1.85     &&  0.27 & 2.81  & 1.21  & 0.33  \\
\noalign{\hrule height 0.5pt}                                                  
$2^+_3\to 0^+_2$ && 206.5  & 22.7   & 23.1 & 91.5    && 253.6 & 29.6  & 114.1   & 27.5  \\
\noalign{\hrule height 0.5pt}                                                  
$2^+_2\to 2^+_1$ && 24.26  & 2.98   & 2.56 & 11.06    && 20.52 & 2.06  & 8.89  & 2.40  \\
\noalign{\hrule height 0.5pt}\end{tabular}
\end{table}
%%%%%%%%%%%%%%%%%%%%%%%%%%%%%%

We analyze the quadrupole transitions of the two nuclei in detail.
In Table \ref{tab:IS2}, we show the quadrupole transition strengths of the isoscalar (IS2), isovector (IV2), proton (P2),
which is equivalent to $E2$, and neutron (N2) parts.
These values are the squared magnitudes of the matrix elements of the transition operators,
defined as $\sum_{i=1}^{Z/N}\sum_m {\cal Y}_{2m}(\bm{r}_i-\bm{r}_{\rm G})^2$ for protons and neutrons,
using the solid spherical harmonics, dividing by the multiplicity of the initial spin.
The similar analysis is performed in the $\alpha+\alpha+n+n$ cluster model of $^{10}$Be \cite{furumoto23}.

For $2^+_1\to 0^+_1$ in $^{10}$Be, the transitions of protons and neutrons give a similar value,
in spite of the larger neutron number than the proton number, and then the IV2 component becomes tiny due to their cancellation.
This indicates that the neutrons in $^{10}$Be are less deformed,
which is consistent with the small quadrupole moment of neutrons in $2^+_1$ of $^{10}$Be, as shown in Table \ref{tab:Q}.
In $^{10}$C, the same tendency can be seen, and the proton transition is similar to and a little bit larger than the neutron transition.
This indicates the small deformation of protons and small quadrupole moment of protons in $2^+_1$ of $^{10}$C.
\red{From this analysis, we can understand the origin of the anomaly of $B(E2)$ in $^{10}$C and $^{10}$Be.}

For other transitions of $2^+_2 \to 0^+_1$, $2^+_3\to 0^+_2$, and $2^+_2\to 2^+_1$,
the proton transitions are smaller than those of neutron in $^{10}$Be and this relation becomes opposite in the mirror nucleus $^{10}$C.
These results are naively reasonable from the number relation of protons and neutrons in the two nuclei.
It is noted that the isovector quadrupole transition is large exceptionally in the transition of $2^+_2 \to 0^+_1$,
and accordingly, the isoscalar transition becomes small.
These results come from the opposite sign of the quadrupole transition matrix elements between the proton and neutron parts.
The same results are reported for $^{10}$Be in Ref.~\cite{furumoto23} and the authors try to investigate the effect of small isoscalar transitions
on the scattering cross section of $^{10}$Be from $0^+_1$ to $2^+_2$ with the $^{12}$C target.
It would be interesting to confirm small transitions in the two nuclei experimentally in the future.

%%%%%%%%%%%%%%%%%%%%%%%%%%%%%% 
\begin{table}[t]
  \caption{
    Magnitude of the quadrupole transition matrix elements for proton ($M_p$) and neutron ($M_n$)
    in $^{10}$Be and $^{10}$C in the multicool calculation. Units are in fm$^2$.
    The values with square brackets are the experimental ones \cite{furuno19}.
  }\vspace*{0.1cm}
  \label{tab:ME}
\renewcommand{\arraystretch}{1.5}
\begin{tabular}{cllllllllllll}
\noalign{\hrule height 0.5pt}
                 &&  \multicolumn{2}{c}{$^{10}$Be} && \multicolumn{2}{c}{$^{10}$C}  \\ \cline{3-4} \cline{6-7}
                 &&  $M_p$      & $M_n$            && $M_p$ & $M_n$  \\
\noalign{\hrule height 0.5pt}
$0^+_1\to 2^+_1$ &&  6.27       &  6.26            &&  6.65             & 6.30 \\
                 &&             &                  &&  [$6.63\pm 0.11$] & [$6.9\pm 0.7 \pm 1.2$]  \\
$0^+_1\to 2^+_2$ &&  1.35       &  3.04            &&  2.46             & 1.29  \\
\noalign{\hrule height 0.5pt}
\end{tabular}
\end{table}
%%%%%%%%%%%%%%%%%%%%%%%%%%%%%%

In Table \ref{tab:ME}, we list the magnitude of the quadrupole transition matrix elements in the two nuclei in units of fm$^2$,
the squared value of which becomes the transition strength.
Taking the absolute square of the matrix elements dividing by the multiplicity of the initial state,
we obtain the same values shown in Tables \ref{tab:IS2}.
We compare the results for $^{10}$C with the experimental values obtained from the inelastic scattering with an $\alpha$ particle \cite{furuno19}.
It is found that our results reproduce well the experiments for $0^+_1\to 2^+_1$
and the components of proton ($M_p$) and neutron ($M_n$) are very similar, as is discussed in Table \ref{tab:IS2}.
For $0^+_1 \to 2^+_2$, the isospin symmetry is confirmed in the exchange of $M_p$ and $M_n$ in the two nuclei, 
which is naively consistent with the number relation of protons and neutrons.

%%%%%%%%%%%%%%%%%%%%%%%%%%%%%% 
\begin{table}[t]
  \caption{
    Values of deformation parameter $\gamma$ of protons and neutrons of $^{10}$Be and $^{10}$C
    extracted in terms of the ratio of the quadrupole transitions of $2^+_2 \to 2^+_1$ to $2^+_2 \to 0^+_1$
    using the Davydov-Filippov rotator model \cite{davydov58}.
    Units of $\gamma$ are in degree.
  }\vspace*{0.1cm}
\label{tab:gamma}
\renewcommand{\arraystretch}{1.5}
\begin{tabular}{crrrrrrrrrrrr}
\noalign{\hrule height 0.5pt}  \vspace*{-0.10cm}
         & \multicolumn{2}{c}{$^{10}$Be} && \multicolumn{2}{c}{$^{10}$C} \\ \cline{2-3} \cline{5-6}
         & Proton   & Neutron  && Proton & Neutron \\
\noalign{\hrule height 0.5pt}                                                                   
Ratio    & 7.01     & 5.98     &&  7.36  & 7.22    \\
$\gamma$ & 21.4     & 20.6     &&  21.6  & 21.5    \\
\noalign{\hrule height 0.5pt}
\end{tabular}
\end{table}
%%%%%%%%%%%%%%%%%%%%%%%%%%%%%%

In $^{10}$Be and $^{10}$C, the bands starting from the $2^+_2$ states dominantly have the $K^\pi=2^+$ component.
Hence, we discuss the triaxiality of the two nuclei using the Davydov-Filippov rotator model \cite{davydov58}.
From the model, we can extract the deformation parameter $\gamma$ according to the following relations in the $E2$ transitions: 
\begin{equation}
  \begin{split}
    \frac{ B(E2;2^+_2 \to 2^+_1)}{B(E2;2^+_2 \to 0^+_1)}
    &= \dfrac{ 20\sin^2(3\gamma) }{7\left( f(\gamma) - (3-2\sin^2 (3\gamma))\sqrt{ f(\gamma)}\right)}
    \\
    f(\gamma)&:=9-8\sin^2(3\gamma)
  \end{split}
  \label{eq:davydov}
\end{equation}
We employ the same formula for the quadrupole transitions of neutrons.

We utilize the quadrupole transitions of $2^+_2 \to 2^+_1$ and $2^+_2 \to 0^+_1$ for protons and neutrons as shown in Table \ref{tab:IS2}.
The parameters $\gamma$ are extracted from the ratios of the two transitions and are shown in Table \ref{tab:gamma}.
It is found that they commonly exhibit values of around 21 degrees for protons and neutrons in the two nuclei.
The $\gamma$ value obtained for the protons of $^{10}$Be is consistent with the 17--22 degrees reported in the molecular orbital model \cite{itagaki02b}.
Based on these results, $^{10}$Be and $^{10}$C are considered to exhibit the triaxiality for protons and neutrons in the ground-state energy region, 
as can be seen in the density distribution shown in Fig.~\ref{fig:density1}.

%%%%%%%%%%%%%%%%%%%%%%%%%%%%%% 
\begin{table}[t]
  \caption{
    Electric quadrupole transitions $B(E2)$ from $4^+$ states to $2^+$ states of $^{10}$Be and $^{10}$C in the multicool calculation
    and the molecular orbital model (MO) \cite{itagaki02b}.
    Units are in $e^2 {\rm fm}^4$.
    In the last column, the ratios of the present results of $^{10}$C to those of $^{10}$Be are shown.
  }\vspace*{0.1cm}
\label{tab:E24}
\renewcommand{\arraystretch}{1.5}
\begin{tabular}{cllllllrrrrr}
\noalign{\hrule height 0.5pt}
                 && \multicolumn{2}{c}{$^{10}$Be} && $^{10}$C && Ratio~\\ \cline{3-4} \cline{6-6} \cline{8-8}
                 && Present   & MO                && Present  && C/Be  \\
\noalign{\hrule height 0.5pt}
$4^+_1\to 2^+_1$ &&     5.89  & 11.1              && 11.29    &&  1.92 \\
$4^+_1\to 2^+_2$ &&     1.27  &                   &&  0.26    &&  0.21 \\
$4^+_1\to 2^+_3$ &&     1.05  &                   && 10.76    && 10.28 \\
\noalign{\hrule height 0.5pt}                               
$4^+_2\to 2^+_1$ &&    0.004  &                   && 0.002    &&  0.46 \\
$4^+_2\to 2^+_2$ &&    0.03   &                   && 0.002    &&  0.66 \\
$4^+_2\to 2^+_3$ &&    31.70  &                   && 150.33   &&  4.74 \\
\noalign{\hrule height 0.5pt}
\end{tabular}
\end{table}
%%%%%%%%%%%%%%%%%%%%%%%%%%%%%%

Finally, we discuss the electric quadrupole transitions from $4^+ \to 2^+$, and the results are shown in Table \ref{tab:E24}.
No experimental data is available, but one theoretical value calculated using molecular orbital model is shown for $^{10}$Be \cite{itagaki02b}.
We also evaluate the ratios of the transitions of $^{10}$C to those of $^{10}$Be.
It is found that three transitions give the ratio more than unity, which is reasonable from the proton number relation.
The remaining three cases give less than unity, but these transitions give small strengths, and it is difficult to obtain the conclusive results from their small ratios.
The transitions of $4^+_2\to 2^+_3$ are very large in the two nuclei because of the large deformation from the linear-chain configurations
of the initial and final states. This result is consistent with the large negative quadrupole moments of two states in the two nuclei, shown in Table \ref{tab:Q}.
  In the transitions, the value of $^{10}$C is very large.
  The large proton radii of the $2^+_3$ and $4^+_2$ states in $^{10}$C, shown in Table \ref{tab:quanta}, can contribute to this enhancement in addition to the large proton number of $^{10}$C.

\section{Summary}\label{sec:summary}

\red{We investigated the quadrupole properties of $^{10}$C and $^{10}$Be and discussed the isospin symmetry in the two nuclei.
This study is motivated by the experiment, in which the electric quadrupole transitions of $2^+_1$ to $0^+_1$ of the two nuclei show similar values
despite their different numbers of protons \cite{mccutchan12}.
This is considered as an anomaly and the same situation occurs in theory.}

In the wave functions of the two nuclei, we generated the optimal configurations using the nuclear model of the AMD.
The AMD configurations are superposed and determined simultaneously to minimize the energy of the total system.
We further optimize the excited-state configurations imposing the orthogonal condition to the ground-state configurations.
This optimization of the multiple AMD configurations is beneficial for describing the various structures of the mean-field and cluster states in a nucleus unifiedly.

In AMD, the nucleon wave function is a Gaussian wave packet and
the centroid parameters of Gaussians are determined in the cooling method for the multiple AMD basis states.
Hence, we call this method the multicool calculation \cite{myo23b}.
In this paper, we applied this method to $^{10}$C and $^{10}$Be and focused on the properties of the quadrupole transitions and moments in the two nuclei.

We can reproduce the energy spectra of $^{10}$C and $^{10}$Be, both of which include elongated linear-chain shapes in the excited states.
We calculate the monopole and quadrupole transitions of protons and neutrons of $^{10}$C and compare the results with those of $^{10}$Be.
For monopole transitions, the good isospin symmetry is almost confirmed;
\red{the exchange of proton and neutron parts of $^{10}$C corresponds to those of $^{10}$Be.}

\red{For quadrupole transitions of $2^+_1$ to $0^+_1$, our model almost reproduces the experimental values for the two nuclei as well as
the predictions of other theories.}
From the quadrupole moments of $2^+_1$, it is found that the protons of $^{10}$C are less deformed than the neutrons,
which is considered to come from the subclosed nature of the $p_{3/2}$ orbit of protons in $^{10}$C.
However, the protons in $^{10}$Be show a large deformation originating from the $\alpha$-$\alpha$ clustering,
which enhances the proton quadrupole moment.
The opposite results on the deformation are obtained for neutrons of the two nuclei,
which indicates the symmetry in the exchange of protons and neutrons in the two nuclei.

We further investigate the quadrupole transitions associated with the other $0^+$ and $2^+$ states in the two nuclei comprehensively,
and evaluate the ratio of $^{10}$C to $^{10}$Be.
It is found that only the transitions of $2^+_1\to 0^+_1$ show the exception and other cases show larger values of $^{10}$C than those of $^{10}$Be,
the latter of which is a reasonable trend from the proton number.
\red{
  The small ratio of the $2^+_1\to 0^+_1$ transitions originates from the small electric quadrupole moment of $2^+_1$ in $^{10}$C,
  which has a $p_{3/2}$ sub-closed nature, as mentioned.
  This is the origin of the anomaly in the electric quadrupole transitions in $^{10}$C and $^{10}$Be.}

We also evaluate the triaxiality in terms of the Davydov-Filippov model with the quadrupole transitions associated with $2^+_2$ in the two nuclei.
The obtained deformation parameter $\gamma$ shows the similar values of around 21 degrees in $^{10}$C and $^{10}$Be to confirm the triaxiality.
For the elongated linear-chain states of the two nuclei, we obtain the large quadrupole transitions between the states of $0^+_2$, $2^+_3$, and $4^+_2$.

The neutron matrix elements of the quadrupole transitions are also discussed together with the proton values. Our results of $0^+_1\to 2^+_1$ in $^{10}$C
agree with the experimental values \cite{furuno19}.
We further provide the values of $0^+_1\to 2^+_2$, which shows the enhancement of the proton part rather than the neutron part in $^{10}$C
and the opposite relation is obtained for $^{10}$Be.

To confirm the theoretical predictions obtained in the present method, future experimental data would be desirable.

\section*{Acknowledgments}
This work was supported by JSPS KAKENHI Grants No. JP21K03544, No. JP22K03643, No. JP25H01268, and JST ERATO Grant No. JPMJER2304, Japan.
This work was also partly supported by the RCNP Collaboration Research Network program as the Project No. COREnet-059.
M.L. acknowledges support from the National Natural Science Foundation of China (Grant No. 12105141),
and the Jiangsu Provincial Natural Science Foundation (Grant No. BK20210277).
Numerical calculations were partly achieved through the use of SQUID at the Cybermedia Center, Osaka University.

\bibliographystyle{apsrev4-2} % aps
\bibliography{reference_c10} % bibtex

\end{document}